\documentstyle[twoside,fleqn,espcrc2]{article}
\title{%
Observing Quantum Tunneling in Perturbation Series%
\thanks{Talk given by H.S. at the 7th Asia Pacific Physics
Conference held between August 19--23, 1997 at Peking, China.}
}
\author{Hiroshi Suzuki and Hirofumi Yasuta\address{%
Department of Physics, Ibaraki University, Mito 310, Japan}}
\begin{document}
\begin{abstract}
It is well-known that the quantum tunneling makes conventional
perturbation series non-Borel summable. We use this fact reversely
and attempt to extract the decay width of the false-vacuum from
the actual perturbation series of the vacuum energy density (vacuum
bubble diagrams). It is confirmed that, at least in quantum
mechanical examples, our proposal provides a complimentary approach
to the the conventional instanton calculus in the strong coupling
region.
\end{abstract}
\maketitle
\section{Introduction}
Usually the quantum tunneling is regarded as a purely
non-perturbative phenomenon. This widely accepted picture emerges
from the observation like this: Consider a simple Hamiltonian of a
metastable system
\begin{equation}
   H={1\over2}p^2+{1\over2}x^2-{1\over4!}gx^4.
\label{one}
\end{equation}
Since the potential energy is not bounded from below, the quantum
system should be defined by the analytic continuation from~$g<0$.
This continuation produces an imaginary part of the quasi-ground
state energy eigenvalue, that is physically interpreted as the total
decay width of the quasi-ground state due to quantum tunneling. In
the WKB (or instanton) approximation, we find
\begin{eqnarray}
   &&{\rm Im}\, E(g)
\nonumber\\
   &&\sim-2\sqrt{3}\left({S_0\over2\pi g}\right)^{1/2}
   e^{-S_0/g}(1+O(g)),
\label{two}
\end{eqnarray}
where $S_0$~is the instanton~\cite{col} action and $S_0=8$ in this
model. Since eq.~(\ref{two}) vanishes to all orders of the expansion
on~$g$, the tunneling effect is invisible in a simple expansion with
respect to the coupling constant. In other words, a simple
(truncated) sum of the conventional Rayleigh-Schr\"odinger
perturbation series of the ground state energy,
\begin{equation}
   E(g)\sim\sum_{n=0}^\infty c_ng^n,
\label{three}
\end{equation}
where $c_0=1/2$, $c_1=-1/32$, $c_2=-7/1536$, $\cdots$, is always
real for~$g$ real. Therefore, one has to utilize a certain
non-perturbative technique to evaluate the tunneling
amplitude~(\ref{two}): This is the conventional wisdom.

However, it is interesting to note that extensive studies started
in the seventies~\cite{rev} have revealed the close relationship
between the tunneling amplitude in the weak coupling
region~(\ref{two}) and the large order behavior of the perturbation
series~(\ref{three}). In fact, it is possible to show
that~\cite{rev},
\begin{equation}
   c_n\sim-{2\sqrt{3}\over\pi(2\pi)^{1/2}}
   \left({1\over S_0}\right)^n\Gamma(n+1/2)(1+O(1/n)).
\label{four}
\end{equation}
In general, at least in super-renormalizable theories, one finds the
following correspondence:
\begin{eqnarray}
   &&\hbox{Tunneling rate in the {\it weak\/} coupling}
\label{five}\\
   &&\Leftrightarrow
     \hbox{{\it Higher\/} order perturbation coefficients}.
\nonumber
\end{eqnarray}
This relation is interesting because the right hand side is hard to
be evaluated, while the left hand side is tractable. The large order
behavior of perturbation series in various systems has been
investigated on the basis of the connection~(\ref{five})~\cite{rev}.

Furthermore it can also be shown that the $O(g)$~correction
in~(\ref{two}) gives rise to the $O(1/n)$~correction for the large
order behavior~(\ref{four}). Therefore, by extrapolating the
relation~(\ref{five}), we expect the following:
\begin{eqnarray}
   &&\hbox{Tunneling rate in the {\it strong\/}
     coupling}
\label{six}\\
   &&\Leftrightarrow
     \hbox{{\it Lower\/} order perturbation coefficients}.
\nonumber
\end{eqnarray}
This possibility seems astonishing, because, although it is very
difficult to systematically evaluate the left hand side, we can
certainly compute the right hand side: This is the basic idea behind
our approach~\cite{our,our2}. (A similar approach was first proposed
in~\cite{kle} on the basis of another kind of resummation method.)
We stress that the higher order corrections to the leading instanton
approximation~(\ref{two}) is difficult: The difficulty is not only
technical but even a principal one---one has to resolve the
``mixing'' between non-trivial configurations and perturbative
fluctuations to avoid the double-counting. Also such an instanton
expansion, i.e., an expansion with respect to~$\exp(-S_0/g)$, is
expected to be asymptotic at best. Therefore if a certain method
can bypass these complications of the instanton calculus,
it will provide a complementary approach to the tunneling phenomenon
in the strong coupling region. This is what we want to propose in
the next section.

\section{Borel resummation of vacuum bubbles}
As the natural extension of~(\ref{one}),
we consider the $O(N)$~symmetric $\phi^4$~model whose action is given
by
\begin{equation}
   S[\phi]=
   \int d^Dx\,\left[{1\over2}\partial_\mu\phi\partial^\mu\phi
   -{1\over2}\phi^2+{1\over4!}g(\phi^2)^2\right],
\label{seven}
\end{equation}
where $\phi^2=\phi\cdot\phi$: Note that the system is metastable
for~$g>0$. We consider the super-renormalizable cases ($D=1$,~$2$,
and~$3$) and assume the appropriate renormalization. Then the
standard instanton calculus~\cite{col} gives the following
expression for the imaginary part of the false-vacuum energy
density,
\begin{eqnarray}
   &&{\rm Im}\,{\cal E}(g)
\nonumber\\
   &&\sim-A_NC_{D,N}\left({S_0\over2\pi g}\right)^{(D+N-1)/2}
   e^{-S_0/g},
\label{eight}
\end{eqnarray}
where $S_0$~denotes the instanton action,
\begin{equation}
   S_0=\cases{8&for $D=1$,\cr
              35.10269&for $D=2$,\cr
              113.38351&for $D=3$,\cr}
\label{nine}
\end{equation}
and $A_N$ and~$C_{D,N}$ are some numbers arising from the Gaussian
integration around the instanton and the collective coordinate
integrations~\cite{bre,our2}.

As noted in Introduction, the leading instanton
approximation~(\ref{eight}) is reliable only for~$g\ll1$. Therefore,
following the basic idea~(\ref{six}), we start with the conventional
perturbative expansion of the vacuum energy density, i.e., a sum of
the vacuum bubble diagrams
\begin{equation}
   {\cal E}(g)\sim\sum_{n=0}^\infty c_ng^n.
\label{ten}
\end{equation}
Once the perturbation series~(\ref{ten}) is obtained, we construct
the Borel-Leroy transform:
\begin{equation}
   B(z)\equiv\sum_{n=0}^\infty{c_n\over\Gamma(n+(D+N)/2)}z^n.
\label{eleven}
\end{equation}
Then we define the vacuum energy density by the Borel integral
\begin{eqnarray}
   &&{\cal E}(g)={1\over g^{(D+N)/2}}\int_0^\infty dz\,e^{-z/g}
   \,z^{(D+N)/2-1}
\nonumber\\
   &&\qquad\times B(z\pm i\varepsilon).
\label{twelve}
\end{eqnarray}
This representation is justified by the Borel summability of
$g<0$~case~\cite{eck} combined with several plausible
assumptions~\cite{our2}.

Now, the fact that all the perturbative coefficients in~(\ref{ten})
(except~$c_0$) have the same signs is important (see (\ref{three})
and~(\ref{four})). The perturbation series of non-alternating sign
is usually called non-Borel summable because the Borel
transform~(\ref{eleven}) develops singularities on the
{\it positive\/} real axis, and thus the integration~(\ref{twelve})
is ill-defined without the~$\varepsilon$-separation. However, these
singularities are welcome for us because these are precisely the
points where the imaginary part emerges---Borel singularities
detect the quantum tunneling. In fact, the leading
instanton approximation~(\ref{eight}) for~$g\ll1$ implies the nearest
singularity from the origin is a square-root branch point
\begin{equation}
   B(z)=-{A_NC_{D,N}S_0^{1/2}\over\sqrt{\pi}(2\pi)^{(D+N-1)/2}}
   (S_0-z)^{-1/2}+\cdots.
\label{thirteen}
\end{equation}
When substituted in~(\ref{twelve}), the branch cut reproduces
the imaginary part~(\ref{eight}) by choosing the upper contour.

Thus, in principle, we construct the Borel transform~(\ref{eleven})
from the actual perturbation coefficients~$c_n$ and substitute it
in~(\ref{twelve}) to extract the ``non-perturbative'' information.
However, this way of working is impossible in practice because the
radius of convergence of the series~(\ref{eleven}) is finite~($=S_0$)
due to the singularity~(\ref{thirteen}). In order to perform the
Borel integration~(\ref{twelve}), we have to analytically continue
the series~(\ref{eleven}) outside the convergence circle: This is
impossible without knowing all the perturbative coefficients.
Fortunately, as is well-known~\cite{rev}, this difficulty can be
avoided by the conformal mapping trick. To apply this trick, the
knowledge on the position of the nearest
singularity~(\ref{thirteen}), i.e., the instanton action~$S_0$, is
important.

In this way, we obtain the~$P$th~order approximation of the imaginary
part~\cite{our,our2},
\begin{eqnarray}
   &&\left[{\rm Im}\,{\cal E}(g)\right]_P
\nonumber\\
   &&=\left({S_0\over g}\right)^{(D+N)/2}
   \int_0^\pi d\theta\,
   \exp\left(-{S_0\over g}{1\over\cos^2\theta/2}\right)
\nonumber\\
   &&\qquad\times
   {\sin\theta/2\over\cos^{D+N+1}\theta/2}
   \sum_{k=0}^Pd_k \sin k\theta,
\label{fourteen}
\end{eqnarray}
where
\begin{eqnarray}
   &&d_k\equiv\sum_{n=0}^k(-1)^{k-n}
\label{add}\\
   &&\qquad\times
   {\Gamma(k+n)(4S_0)^n\over(k-n)!\,\Gamma(2n)\Gamma(n+(D+N)/2)}
   c_n.
\nonumber
\end{eqnarray}
Note that eq.~(\ref{fourteen}) is {\it solely\/} expressed by the
first $P$~perturbative coefficients~$c_n$ and the value of the
instanton action~$S_0$~(\ref{nine}). In this integration, the
contribution around the origin~$\theta=0$ is proportional
to~$\exp(-S_0/g)$, the leading instanton behavior. Other parts might
be regarded as the higher order corrections to it: This is true at
least in quantum mechanical cases~\cite{our,our2}.

\section{Numerical tests}

The validity of our formula~(\ref{fourteen}) has been tested
numerically in quantum mechanics~$D=1$ and $D=2$, tunneling on
line~\cite{our,our2}. For~$D=1$, perturbative coefficients of
the vacuum energy, i.e., the ground state energy, to very high orders
are available. The first several coefficients are
\begin{eqnarray}
   &&c_0={N\over2},\quad c_1=-{N(N+2)\over 96},
\nonumber\\
   &&c_2=-{N(N+2)(2N+5)\over4608},
\label{fifteen}\\
   &&c_3=-{N(N+2)(8N^2+43N+60)\over221184},\quad\cdots.
\nonumber
\end{eqnarray}
With the aid of computer, it is not difficult to compute~$c_n$ to,
say,~$n=50$. The exact complex quasi-ground state energy is also
available by a numerical diagonalization of the Hamiltonian.
Therefore, we can compare our formula~(\ref{fourteen}) and the
instanton result~(\ref{four}) with the exact value of the imaginary
part. For the detail, we refer the reader to Refs.~\cite{our,our2}.
For~$D=1$, we see an excellent convergence of~(\ref{fourteen}) to the
exact value in a whole range of the coupling constant; the
proposal in fact gives rise to the improvement of instanton calculus,
as we announced. For example, for~$g=4$, which belongs to the strong
coupling region~\cite{our2}, eq.~(\ref{fourteen})~with~$P=5$ (thus
only the first five perturbative coefficients!) gives only a few
percent error, while the leading instanton approximation~(\ref{eight})
is about two times larger than the correct value.

For~$D=2$, the situation is yet not obvious. We have
calculated vacuum bubble diagrams to five loop orders under the same
renormalization condition assumed in~(\ref{eight}). We
found (with an appropriate shift of the origin of the vacuum
energy)~\cite{our2}:
\begin{eqnarray}
   &&c_0=0,\quad c_1=0,
\nonumber\\
   &&c_2=-{N(N+2)\over3}\times8.833\times10^{-5},
\label{sixteen}\\
   &&c_3=-{N(N+2)(N+8)\over27}\times3.012\times10^{-6},
\nonumber
\end{eqnarray}
and
\begin{eqnarray}
   &&c_4
\nonumber\\
   &&=-{N(N+2)(N^2+6N+20)\over81}\times5.657\times10^{-8}
\nonumber\\
   &&\quad-{N(N+2)^2\over9}\times1.006\times10^{-7}
\label{twentyone}\\
   &&\quad-{N(N+2)(5N+22)\over81}\times2\times10^{-7}.
\nonumber
\end{eqnarray}
Unfortunately, any converging behavior could not be observed and
it was impossible to draw a definite conclusion on the
convergence of~(\ref{fourteen})~\cite{our2}. It is not clear whether
this is due to the lack of orders of the perturbation series, or
there exists a fundamental obstruction for our approach which we did
not encounter in quantum mechanics. To clarify this point, much
higher order perturbative calculation is certainly desirable.

\section{Discussion}
We have proposed a new approach to the tunneling phenomenon in
super-renormalizable field theories. Our approach utilizes only the
information of the conventional perturbation series around the naive
vacuum~$\phi=0$ (and the value of the instanton action). We have
verified numerically that, at least in quantum mechanical
cases~($D=1$), we can extract a very accurate tunneling rate from the
perturbation series. The procedure thus provides a complimentary
approach to the instanton calculus in the strong coupling region.
In~$D=2$, unfortunately, the number of orders of the perturbation
series we computed is yet insufficient to draw a definite conclusion
on the convergence.

As another test of our proposal, one of us~\cite{yas} recently
applied it to the Gaussian propagator model, in which the
perturbative calculations in $D=1$,~$2$, $3$ and~$4$, up to the
ninth order of the loop expansion are known~\cite{ber}. It was
found that eq.~(\ref{fourteen}) rapidly converges in the whole range
of the coupling constant for lower dimensional cases, $D=1$ and~$2$,
and, at least in the strong coupling region for higher dimensional
cases, $D=3$ and~$4$.

As a general question,
one might ask ``To which extent we can expect the applicability of
such a perturbative approach to tunneling phenomenon?'' For example,
we see that our approach {\it cannot\/} work for the double-well
potential problem: The perturbation series around one of potential
minima is non-Borel summable~\cite{rev} and thus the Borel integral
produces the imaginary part. However this imaginary part must be
fake because the potential energy is bounded from below (i.e., no
decaying process in involved). This fake imaginary part is believed
to be canceled by the multi-instanton contributions~\cite{rev}. In
other words, in this system, the information of the perturbation
series around the naive vacuum is not sufficient to specify the
original physical quantities. Therefore the natural answer to the
above question is: Such a perturbative approach to tunneling
phenomena is expected to be applicable {\it if the system can be
defined by an analytic continuation of another system in which the
Borel summability (of the energy density) is guaranteed}.
Eq.~(\ref{seven}) with $D=1$,~$2$ and~$3$ is expected to be precisely
such a system.

Stated differently, for our approach to be workable, it is important
that there is no non-trivial topological sector in the configuration
space. The instanton in the system~(\ref{seven}) is
``non-topological''---the so-called bounce solution---and it can mix
with the trivial perturbative sector. Our approach systematically
counts the contribution of the perturbative fluctuations while
avoiding the double counting of the bounce configuration. If there
exist a non-trivial topological sector which does not mix with the
trivial perturbative sector, such as the instanton in the double well
potential, we will have to supplement the ``truly'' non-perturbative
information. This is the principal limitation of our approach.

When one considers a renormalizable field theory, such
as~(\ref{seven}) with~$D=4$, new difficulty may arise~\cite{our2}.
The renormalon~\cite{rev}---another known source of the Borel
singularity---emerges in general. The renormalon has two effects:
i)~The contribution of renormalons to the Borel singularity produces
new imaginary part besides the quantum tunneling. ii)~The UV
renormalon suggests the triviality of the model. In~(\ref{seven})
with~$D=4$, when~$g>0$, there is no renormalon singularity on the
positive real $z$~axis, because the model is asymptotically free.
However the model with~$g<0$ has the UV renormalon and is expected
to be trivial. Therefore the meaning of the {\it true\/} tunneling
amplitude, which should be defined by the analytic continuation
from~$g<0$, is not obvious. Presumably, in renormalizable field
theories, our proposal is workable only with an UV cutoff.

Finally, we would like to comment that our approach is applicable
not only to the quantum tunneling problem, but also to metastable
problems in statistical mechanics. For example, the free energy of
the Ising model below the critical temperature acquires the
imaginary part when it is analytically continued from the positive~$H$
to the negative~$H$ ($H$ is the external magnetic field and we
assumed the Ising spins are originally aligned to the positive
direction). The imaginary part might be physically interpreted as the
inverse of the relaxation time of the metastable state. Our approach
is applicable to the perturbation series on~$H$, which can be
computed from the low temperature expansion of the free energy.
Then we expect the validity of our approach in the strong~$H$
which will provide a complimentary approach to the conventional
droplet calculation. A study along this line is in progress.

This work was supported in part by the Ministry of Education
Grant-in-Aid for Scientific Research, Nos.~08240207, 08640347,
and 08640348.

\end{document}